# pch2csd: an application for converting Nord Modular G2 patches into Csound code

**Gleb Rogozinsky**
The Bonch-Bruevich
Saint-Petersburg State
University of Telecommunications,
St. Petersburg, Russia
`gleb.rogozinsky@gmail.com`

**Mihail Chesnokov**
JSC SEC "Nuclear Physics Research"
St. Peterburg, Russia
`chesnokov.inc@gmail.com`

**Eugene Cherny**
Åbo Akademi University, Turku, Finland
ITMO University,
St. Peterburg, Russia
`eugene.cherny@oscii.ru`

## ABSTRACT

The paper presents the pch2csd project, focused on converting patches of popular Clavia Nord Modular G2 synthesizer into code of Csound language. Now discontinued, Nord Modular G2 left a lot of interesting patches for sound synthesis and algorithmic composition. To give this heritage a new life, we created our project with the hope for being able to simulate the original sound and behavior of Nord Modular.

## 1. INTRODUCTION

Clavia Nord Modular was one of most remarkable hardware synthesizers of the late 90s. It inspired a whole new generation of modular synthesizer fans by combining the immediacy of dedicated hardware with the power and flexibility of computer-based programming. The suite of components comprising the NM project included an extensive list of modules (from oscillators to effects), an attractive software graphical interface and on the hardware side, keyboard and rack options. This synthesizer has been used extensively by many artists such as Astral Projection, Autechre, The Chemical Brothers, Somatic Responses, Junkie XL, Mouse on Mars, Nine Inch Nails and Covenant among others.

What really makes Nord Modular unique and sustains its relevance is the active community, which created a rich archive of patches. Alongside their inherent use to musicians, these patches can also incorporate a valuable educational aspect, inspiring people to develop their own creative signal processing and sequencing skills.

Unfortunately, Clavia ceased the Nord Modular production in 2009, and at present time it could be hard to buy the synth from the second-hand market to explore the vast collection of creative patches that were made by the community since the late 90s.

To facilitate the liberation of the NM patches from the closed source software and provide it with a continuing existence, we started a project called pch2csd in 2015 to (re)implement the Clavia Nord Modular G2 sound engine in Csound, a well-known sound and music computing system. Csound is one of the oldest computer music systems, it carries the sounds (and patches) from the past, which were originally compiled on mainframe computers back in the days, but have been continually re-compiled throughout the intervening years. Even now Csound can compile many sources written for the older versions of the language in the 80s. But despite it's heritage, at the present time Csound is being actively developed, and currently, the Csound code can interoperate with almost any programming language, providing robust, high performance and cross-platform solution for sound computing in real-time applications. Csound had been ported to iOS, Android and Raspberry Pi, there are Csound-based DAWs, tools to transform orchestra files to VST/VSTi plugins, a library for the Unity3D game engine [1], Jupyter notebook bindings and many more [1].

The fist report on the pch2csd project was made at the Third International Csound Conference (St. Petersburg, Russia 2015) [2]. We presented the main concept of the project, the brief description of the code and a very simple "proof of a concept" working example. Since then, a number of improvements has been made, including new modules, mapping tables, and other code generation improvements.

This paper provides the report on current project status, as well as the engine implementation details.

The paper is organized as follows. Section 2 lists some other works on simulating music hardware in Csound. Section 3 provides the overview of Nord Modular G2 and it's patch format. Section 4 describes the implementation of the sound engine in Csound language, including the transformation from the NMG2 patch format. Section 5 discusses several patch conversion examples. Section 6 describes the limitations of the project. In the section 7 we conclude the paper and provide some directions for future developments.

## 2. RELATED WORK

Csound already has its own history of hardware emulation. There are several opcodes which emulate Moog filters and also the mini-Moog synthesiser. The DirectHam-mond synthesiser written by Josep Comajuncosas emulates Hammond with the Leslie effect [3]. The Csound Book gives a simulation of the famous Roland TB303 [4]. Another example of TB303 was given by Iain McCurdy, who also wrote a code for emulation of Korg Mini Pops 7 and Roland TR-808 drum modules [5]. Csound had been used for emulating Yamaha DX7 [6]. Several experiments were carried to emulate the sound of Access Virus and Roland JP80xx, for example [7]. Eugenio Giordani and Alessandro Petrolati created a commercial app which emulated VCS3 by EMS [8].

---

[1] A (not extensive) list of tools and instruments built on top of Csound can be found at the community websie, URL: `http://csound.github.io/create.html`

The DX7 implementation stands apart from other examples, as it translates the synth's presets into a dynamically generated Csound code. Essentially the same approach is used for this project, but the code generation is more complex in our simulation.

## 3. NORD MODULAR OVERVIEW

Nord Modular and Nord Modular G2 were a unique hybrid software-hardware systems. They ran on several DSPs, but were controlled through a GUI using the software editor which is still available from the Clavia website [2]. Once (re)programmed, device can be run in a stand-alone hardware-only mode. For programming, the device should be connected to PC or Mac via USB cable. One can also find a freely downloadable copy of demo editor.

The Nord Modular system works only in real-time so the number of modules and voices are the most important parameter. Different modules cause different load. To optimize the performance several modules were used with almost identical functions, but with different operability. For example, *OscA* oscillator allows changing its wavetype on a flight, comparing to *OscD*, which uses the same wavetypes but demands the patch to be recompiled. There are two main parts of the patch: *voice* and *fx*. The *voice* part is a subject of polyphony, comparing to *fx* part which receives the mix of all voices. Obviously, the greater the number of voices played simultaneously, the lesser number of modules can be used before overrun.

Nord Modular G2 uses fixed sampling rate of 96 kHz for audio signals and 24 kHz for control rate signals (i.e. modulation). The overall number of modules is around 200, including numerous sound generators, envelopes, lfo, filters, random generators, mixers, delay units, switches, MIDI units, logics, waveshapers, sequencers and FXs.

There are 17 sound generators, most of which provide classical waveshapes with modulations. There are also noise generators, simple physical models and a built-in DX7 model. The Filter part consists of 14 filters, i.e. various LP, HP, BP and BR models, and also comb filter, formant filter and a 16-band vocoder. The section also includes three different equalizers of up to 3 bands. The Envelope section provides 9 envelope generators from simple Decay-only generator to multistage one. There are 16 units in a Mixer sections starting from one channel volume controller to 8 channel mixing unit. The FX part includes chorus, phaser, flanger, digitizer, frequency and pitch shifters, scratcher, reverb and a compressor. Delay units are included in the separate Delay section. There are 10 different units, from static one-tap delay to complex stereo delays. The Switch section contains various switches and dmux/mux units. The Logic includes common binary logic units, i.e. typical logical functions, flip-flops, pulse dividers and counters. There are also ADC and DAC converters. All waveshaping units are placed in the Shaper part. There are 7 of them, i.e. clip, drive, saturation, rectifier and wrapper units. The Level section relates to any modulation types. It also contains the envelope follower unit. The Random section includes 6 various random generators. The Sequencer part includes 5 sequencers of different kinds. All of sequencers can be linked in chain. There also MIDI, In/Out and Note sections. The detailed info on each unit can be found in official NM2 manual.

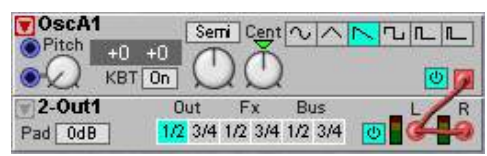

Figure 1. An example of the input-to-input connection: the output of the oscillator is connected to the both (left and right) inputs of the stereo output module.

A sound programmer connects the modules using virtual cables. The main cables types are: red (audio signals @ 96 kHz), blue (control signals @ 24 kHz), yellow (logic signals in the form of pulses @ 24 kHz), orange (logic signals @ 96 kHz).

In addition to these cable types there are also two user-defined types: green and violet. Actually the user can color any cable in any color using the context menu of an editor, i.e. change color of red cable into yellow. Also there are 'dead' cables of grey color. They appear if you break the valid link of modules.

Another peculiar feature of Nord Modular system is an ability to connect one input to another. So if you want to connect the output of a generator to both inputs of the *Out* module, it is allowed to connect one input to another (Fig. 1), and the output from the generator will be passed to both inputs.

Also, several modules can change their type, i.e. from red to blue, depending on input connections, i.e. connecting a red cable to any input of mixer turns it from default blue type to a red one.

Moreover, there are some hidden modules which were not included in the release version of the editor, i.e. Resonator, Driver, AR-Env, PolarFade. Most of them are not working, but some can be useful [3].

### 3.1 Patch format

Nord Modular G2 stores patches in a binary file with the extension *pch2*. The format is proprietary and has not been officially published by Clavia, so we rely on a community effort to decode the format done by Michael Dewberry in the 2005 [9]. According to the information he published, the *pch2* file starts with a text header terminated by the NULL byte. It contains general information about the patch (e.g. version, file type, etc.) in human readable form. This text header is followed by a two-byte binary header, representing the patch format version and the file type (a *patch* or a *performance*). All other information is stored in a series of "data objects". A byte layout of some objects is presented in the Appendix. For the purpose of this project we do not use such objects as Knob Assignments, MIDI Controller Assignments, Module Names and Textpad. Hence we read only those, representing the actual sound modules and their connections. For this, we first read Patch Description object (Table 3), then read two Module List objects (Table 4) for VA and FX [4], then skip the Mystery Object [5] (Table 1), then read two Cable List objects for VA and FX (Table 2), and finally we read two Module Parameters objects for VA and FX (Table 5).

---

[2] Nord Modular G2 official page, URL: http://www.nordkeyboards.com/downloads/legacy/nord-modular-g2

[3] A discussion about hidden modules in Nord Modular G2: http://www.electro-music.com/forum/viewtopic.php?t=54650

## 4. IMPLEMENTATION

### 4.1 Overview

The project is implemented as a C program. The program extracts module and connection lists, as well as parameter values from the *pch2* file, and then composes a *csd* file from Csound code templates and mapping tables stored as plain-text files. We currently have 100 modules implemented as Csound user-defined opcodes (including filters, mixers, etc.) and 35 hand-crafted mapping tables to map values from the linear MIDI range to non-linear parameter range. Storing the data as text allows anyone to change the behavior of the modules without touching the C code.

### 4.2 Modeling NMG2 modules

We began with the challenges related to oscillators modeling. Any sound synthesis algorithm starts from some generator, thus the first and the most important part of the patch to be simulated is an oscillator section. The Nord oscillators produce aliased waveforms at 96 kHz. The Figure 2 shows the amplitude spectra of Clavia's sawtooth. The audio was recorded at 96 kHz/24 bit on the Focusrite Saffire PRO 40 audio interface. The solid line at the right border of plot is a Nyquist frequency (48 kHz). The spectra was calculated using 2048 points FFT with Hann window. We can clearly see the aliasing part of the spectra, mirrored from the Nyquist frequency. This feature of Nord Modular distinguishes it from the popular family of so-called analog-modeling synthesizers, which typically produce alias-free waveforms, and makes it possible to simulate the corresponding waves by simple generation of ideal piece-wise functions. Numerous records of the oscillator waveforms also prove it well.

---

[4] Note, that objects like this appear twice in the patch file, in the Voice Area and in the FX Area.
[5] We do not currently know the purpose of this object, but in our project we are only interested in a list of modules and their connections to reconstruct patches in Csound, and we have not yet found any cases where we would need to inspect this object more closely.

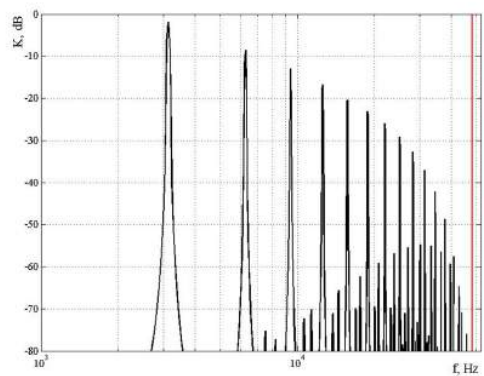

Figure 2. An amplitude spectra demonstrates the aliasing in the NMG2 sawtooth oscillator waveform. A solid vertical line to the right is the Nyquist frequency.

Filter analysis has been performed through connecting white noise generator to the corresponding filter of Clavia and comparing the output amplitude response with the filter models created by the authors.

### 4.3 Csound implementation details

To simulate the sound and behavior of modules of Clavia Nord Modular G2 we used Csound [10]. It is an open-source language for computer music and digital audio processing developed in MIT 1986 by Barry Vercoe and recently developed by a global community of enthusiasts. Csound can be run on all popular platforms including Android and iOS. Through Csound API it can be embedded in lots of applications, which makes it a nice candidate for Nord simulation together with a great amount of included opcodes [6].

Csound document typically consists of two sections. The *Instruments* section contains all definitions of instruments to be used, and the *Score* part contains timeline events and function tables. Each of Nord modules is modeled as an User-Defined Opcode (aka UDO). At the conversion from pch2 to Csound the UDOs, which relate to patch's modules are read from txt files and included to the Csound document. Csound compiler reads code lines from top to the bottom, which is completely different approach to graphical system of patching, like MAX, Pd or Nord. Fortunately, Csound has a special patching system called *zakspace*. It provides a given number of audio rate (a-rate) and control rate (k-rate) buses to intercommunicate between different instruments and UDOs. There are separate commutation matrix for audio rate signals and another one for control rate signals. They are completely independent from each other, comparing to Nord patching system, where the cables are sequentially numbered. A separate part of our code solves that disparity by renumbering the cables after reading their numbers from the patch file. Comparing to typical behavior of Csound opcodes, where each opcode typically accepts an input data as some parameters and outputs the result, our zak-based UDOs do not have any outputs.

---

[6] All up-to-date Csound features can be found on the community website: http://csound.github.io

In more practical view, the typical Csound opcode/UDO is applied like this:

```
aOut1 SomeOpcode aIn1, aP1, kP2 [, ...]
```

where *aIn1* is an input of audio type, *aP1* is an a-rate parameter, *kP2* is a k-rate parameter, and *aOut1* is an a-rate output.

In our system we have

```
SomeOpcode aP1, kP2 [, ...], kIn1, kOut1
```

where *aP1* is an a-rate parameter, *kP2* is a k-rate parameter, *kIn1* is a number of some k- or a-rate bus to read input data from, and *kOut1* is a number of some k- or a-rate bus to send data to.

After parameter field our UDOs have IO field, in which the numbers of buses in zak space are listed. Using described approach we are free to list the opcodes in any order, just like Nord Modular user can add the modules in arbitrary order. So the first indexed module can be easily the last one in the audio chain.

Another important aspect to be described here is mapping. Nord modules have several different controllers of a lot of ranges, i.e. audio frequency range, amplitude range, normalized values in the range from 0 to 1, delay time ranges, envelope stage durations, etc. The real values of the controllers can be seen only when using editor. The patch file stores 7 bit MIDI values without any reference to appropriate range. It made us to manually fill the table with data types, ranges and values. Special mapping files of pch2csd project contain numbers of tables to be used for each parameter of each module.

I.e Module #112 (*LevAdd* module) table contains following lines:

```
s 2 LVLpos LVLlev
d BUT002
```

It means that the mapping table for the first parameter of *LevAdd* (value to add to input) is dependent on a second (*2*) parameter, which is the two-state button (table *BUT002*). The button switches *LevAdd* from unipolar to bipolar range of values. The mapping tables are placed in a separate subdirectory of a project.

Also during the development we had to solve a polymorphism issue. Several Clavia modules are polymorphous. Unfortunately there is no direct indication of current module type (a-rate or k-rate). It can be discovered only through analyzing cable connections. So our algorithm checks the module type, and its input connections. In case of non-default type of the input, the corresponding module twin is used instead of the default one.

## 5. EXAMPLES

Here we demonstrate the conversion of a real Clavia's patch (see picture ...). It consists of a noise generator, a simple 1-pole LP filter and an Output module. We give this extremely primitive patch not for the purposes of timbre discussion but rather to present a clear example of how our converter works.

The converter default output file is *test.csd* in the program directory. Comments are given after semicolons.

The algorithm detects three different types of modules and take their UDO definitions from the library.

```
sr    = 96000  ; audio sampling rate
ksmps = 16     ; times k-rate lower than a-rate
nchnls = 2     ; number of output channels
0dbfs = 1.0    ; relative amplitude level

zakinit 4, 3 ; Init zak-space

opcode Noise, 0, kkk ;White Noise generator
kColor, kMute, kOut  xin
if kMute!=0 goto Mute ;mutes sound if Mute is On
aout rand 0.5, 0.1 ; seed value 0.1
aout tone aout, kColor ; Csound simple LP filter
zaw aout, kOut
Mute:
endop

opcode FltLP, 0, kikkkkk ;One-pole LP filter
; Keyboard Tracking has not been implemented yet
kKBT, iOrder, kMod, kCF, kIn, kModIn, kOut xin
ain zar kIn
kmod zkr kModIn
aout tonex ain, kCF+kmod*kMod, iOrder
zaw aout, kOut
endop

opcode Out2, 0, kkkkk; Output module
; Only stereo output has been implemented
kTarget, kMute, kPad, kL, kR xin
if kMute!=0 goto Mute:
aL zar kL
aR zar kR
outs aL*kPad, aR*kPad
Mute:
endop

opcode Constant, 0, kk   ;Constant value
    ; Only bipolar mode now
kVal xin
zaw kVal, kOut ; CHANGE
endop

instr 1; VA section
    Noise 10000,0,2
    FltLP 0,1,0.5,1050,2,2,3
    Constant 24,2
    Out2 0,0,1,3,0
endin

instr 2; FX section
endin

; Here goes the score part of Csound.
; We just let it run for a long time...
i1 0 [60*60*24*7]
i2 0 [60*60*24*7]
```

We use two special buses per each matrix. Buses #0 contain silence. They connected to inputs without cables. Buses #1 are trash collectors. If output is not connected, it goes there.

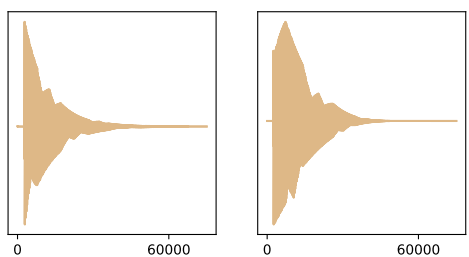

Figure 3. Sawtooth waveform's amplitude modulated with an envelope. Nord Modular G2 is on the left, our implementation is on the right.

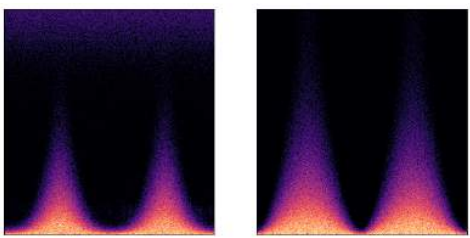

Figure 4. White noise filtered with LFO modulated low-pass filter. The file lengths are equal to 3 seconds. Nord Modular G2 (left) compared with our implementation (right). The Clavia's LFO is not exactly a sine.

## 6. EVALUATION

The practical study of the software shows several issues. Beginning from the oscillator block of the synthesis process, we discovered that Clavia oscillators run at arbitrary initial phase. So at each run or patch re-compilation the composite waveform of the sum of oscillators slightly differs. Also the exponential form of Clavia envelope generators seems to be sharper than real exp function 3. We also discovered that typical sine LFOs actually produce a bit asymmetric function. The figure 4 shows the difference between spectrograms of signals at the output of modulated LPF.

During developing we estimated several main limitations. First is related to the modules with non-determinate behaviour, i.e. all modules from the Random group. Each of those modules generates a random sequence, in which the current value depends on previous one. Such a behavior is not typical for Csound random generators. Another limitation relates to the patch presets. Each of Nord Modular patch contains 8 presets called Variations. User can select the desired preset from the interface. This feature has not been implemented yet. Despite listed limitations, we hope to completely overcome them in our future work.

Certainly, the in-depth evaluation of such project should include much more aspects than listed. Meanwhile, at the present stage of developing we are concentrated mostly on creating simulation which should be close to the original sound of NM2.

## 7. CONCLUSIONS AND FUTURE WORK

At its present state, the pch2csd project is ready for further development by international computer music enthusiasts. The core has been test to work on Windows and OSX systems. The user is able to open Clavia's patch file format pch2. The conversion log shows status of all dependencies, i.e. Csound UDOs, mapping tables, etc. It allows user to update the dependencies and also create his or her own versions of UDOs and mappings if needed. The module completion status is far from uniform. Most of straightforward modules are finished, i.e. mixing operations or switching. Random section seems most difficult, because of its ambiguous behavior. The same should be reported on FX section, although the actual quality of the reverb or chorus modules does not correlate with algorithmic patch behavior, which makes it not so critical.

Another important goal we started working on recently is to make the project hackable, so users would be able to easily modify module implementations to contribute to the project or to modify sound for their own tastes. The plain text modules and mapping tables were made as well as for that purpose, but the user experience here is still poor. As a first step to mitigate this we plan to develop a simple Electron-based UI [7] with an embedded text editor in the near future.

Our next to-do after providing a completely working solution is an integration with some existing Clavia patch editor. It will actually establish the new Clavia-based software modular system running on a Csound core. Also, the native Csound developments, i.e. Cabbage Studio [8] (a graphical UI for Csound build by Rory Walsh) seem very promising in the context of further integration.

Current tool sources can be found on the GitHub [9]: `https://github.com/gleb812/pch2csd`.

**Acknowledgments**

This work has been partially financially supported by the Government of the Russian Federation, Grant #074-U01.

---

[7] URL: `https://electron.atom.io/`
[8] URL: `https://csound.github.io/frontends.html#cabbage-and-cabbage-studio`
[9] The code, as well as sound examples, mentioned in the previous section, are also preserved at Zenodo, URL: `https://zenodo.org/record/581204`

## Appendix: NMG2 binary patch format

| Field name  | Value / Comment | Bits |
|---|---|---|
| Header byte | 0x69 | 8 |
| Length      | 0x4a | 16 |
| *Unknown*   |      |    |

Table 1. Mystery object

| Field name  | Value / Comment | Bits |
|---|---|---|
| Header byte | 0x52 | 8 |
| Location    | 0 / 1 : FX / Voice | 2 |
| Unknown     | - | 14 |
| Cable count | Nr. of cables in the area | 8 |
| *Then, for each cable according to cable count:* | | |
| Color       | 0-6 | 3 |
| Module from | -   | 8 |
| Jack from   | -   | 6 |
| Type        | 0: in-to-in, 1: out-to-in | 1 |
| Module to   | -   | 8 |
| Jack to     | -   | 6 |
| *End of iteration* | | |

Table 2. Cable list binary encoding

| Field name | Value / Comment | Bits |
|---|---|---|
| Header byte | 0x21 | 8 |
| Length | - | 16 |
| Unknown | - | 12 |
| Voice Count | - | 5 |
| Height of FX/VA bar | - | 14 |
| Unknown | - | 3 |
| Red cable visibility | 0: off, 1: on | 1 |
| Blue cable visibility | 0: off, 1: on | 1 |
| Yellow cable visibility | 0: off, 1: on | 1 |
| Orange cable visibility | 0: off, 1: on | 1 |
| Green cable visibility | 0: off, 1: on | 1 |
| Purple cable visibility | 0: off, 1: on | 1 |
| White cable visibility | 0: off, 1: on | 1 |
| Mono/Poly | - | 2 |
| Active variation | 0-7 | 8 |
| Category | 0: No Cat, 1: Acoustic, 2: Sequencer, 3: Bass, 4: Classic, 5: Drum, 6: Fantasy, 7: FX, 8: Lead, 9: Organ, 10: Pad, 11: Piano, 12: Synth, 13: Audio In, 14: User 1, 15: User 2 | 8 |
| Padding | | - |

Table 3. Patch description binary encoding

| Field name | Value / Comment | Bits |
|---|---|---|
| Header byte | 0x4a | 8 |
| Location | 0: FX Area, 1: Voice Area | 2 |
| Module Count | Nr. of modules in the area | 8 |
| *Then, for each module according to module count:* | | |
| Module type | | 8 |
| Module index | Module ID | 8 |
| Horiz. position | - | 7 |
| Vert. position | - | 7 |
| Color | - | 8 |
| Appendix | - | 4 |
| *If Appendix != 0:* | | |
| Unknown | - | 2 |
| Hidden parameter | - | 4 |
| *End of iteration* | | |
| Padding | | - |

Table 4. Module list binary encoding

| Field name | Value / Comment | Bits |
|---|---|---|
| Header byte | 0x52 | 8 |
| Length | | 16 |
| Location | 0 / 1: FX / Voice | 2 |
| Module count | - | 8 |
| *For each module:* | | |
| Module index | - | 8 |
| Param. count | Nr. of parameters | 8 |
| *For each parameter:* | | |
| Variation | 0 | 8 |
| Value | 7-bit MIDI value | 7 |
| Variation | 1 | 8 |
| ... | | |
| Variation | 8 | 8 |
| Value | 7-bit MIDI value | 7 |
| *End of both iterations* | | |
| Padding | 8 | 8 |

Table 5. Module parameters binary encoding